\documentclass[onecolumn,showkeys,showpacs,superscriptaddress,notitlepage]{revtex4-1}
\usepackage{graphicx}
\usepackage{amsmath,amsfonts,amssymb}
\usepackage{morefloats}
\usepackage{color}

\newcommand{\noi}{\noindent}
\newcommand{\be}{\begin{equation}}
\newcommand{\ee}{\end{equation}}

\begin{document}
\title[ ]{The Peace Mediator effect: Heterogeneous agents can foster consensus in continuous opinion models}

\author{Daniele Vilone}\email{daniele.vilone@gmail.com}
\affiliation{LABSS (Laboratory of Agent Based Social Simulation),
Institute of Cognitive Science and Technology,
National Research Council (CNR), 
Via Palestro 32, 00185 Rome, Italy}
\author{Timoteo Carletti}\email{timoteo.carletti@unamur.be}
\affiliation{naXys, Namur Center for Complex Systems, University of Namur, rempart de la Vierge 8, B 5000 Namur, Belgium}
\author{Franco Bagnoli}\email{franco.bagnoli@unifi.it}
\affiliation{Department of Energy, Physics of Complex Systems Laboratory, University of Florence
\& National Institute of Nuclear Physics (INFN). Via S. Marta, 3. 50139 Florence, Italy.
Centre for the Study of Complex Dynamics (CSDC). Via Sansone, 1. 50019 Sesto F.no, Florence, Italy}
\author{Andrea Guazzini}\email{andrea.guazzini@unifi.it}
\affiliation
{Department of Science of Education and Psychology, \& Centre for the Study of Complex Dynamics (CSDC). University of Florence.
Via di San Salvi, 12, Building 26, 50135 Florence, Italy}

\maketitle

\ 

\

\

\section*{Abstract}
Statistical mechanics has proven to be able to capture the
fundamental rules underlying phenomena of social aggregation and
opinion dynamics, well studied in disciplines like sociology and
psychology. This approach is based on the underlying paradigm that
the interesting dynamics of multi-agent systems emerge from the
correct definition of few parameters governing the evolution of each
individual. Into this context, we propose a particular model of opinion
dynamics based on the psychological construct named "cognitive
dissonance". Our system is made of interacting individuals, the
agents, each bearing only two dynamical variables (respectively
``opinion'' and ``affinity'') self-consistently adjusted during time
evolution. We also define two special classes of interacting
entities, both acting for a peace mediation process but via
different course of action: ``\textit{diplomats}'' and
``\textit{auctoritates}''. The behavior of the system with and
without peace mediators ($PM$s) is investigated and discussed with
reference to corresponding psychological and social implications.

\ 

\noindent \textbf{Keywords:} {opinion dynamics, complex systems, peace mediation, social psychology, cognitive dissonance} \\
\\
\textbf{PACS:} {89.65.Ef, 89.75.-k, 82.20.Wt}

%{\bf Word count: } 2600.

\newpage

\section{Introduction}

In recent years we have seen the emergence of a new breed of
professionals broadly called Peace Mediators, $PM$s for short,
involved in the process of peace (re)construction. They are usually deployed
in countries torn by conflict or post-conflict areas in order to
create conditions for sustainable peace. $PM$s actions aim to
reduce the fragmentation among different parts of the society until
a widespread consensus is achieved and peace can be maintained.

Our model is based on the assumption that it is possible to study
the evolution of a social phenomenon directly by considering a few
attributes of the individuals coupled by specific interaction
rules. For these reasons, we adopt an agent based model, in which
local rules are inspired by the \textit{cognitive dissonance}~\cite{fes58},
a cognitive construct that rules the evolution of human
social cognition~\cite{bag07}. According to the Cognitive Dissonance Theory,
when unknown individuals interact, they experiment an
\textit{internal conflicting state} because of their reciprocal lack
of information. In order to avoid the cognitive dissonance,
individuals adopt heuristics strategies with the aid of \textit{mental
schemes}~\cite{2}, that is, symbolic and synthetic representations built up
through inferential, imaginative and emotional processes. Because
mental schemes can be upgraded in real time during interactions with
other individuals, they are utilized as a guidance for quick decisions
in stereotypical situations. For instance, the mutual affinity is the
mental scheme employed by agents to overcome the lack of information
about the others (that is, the cognitive dissonance) and to perform the
optimal choice in terms of opinion production. In particular, two heuristic
strategies are mainly employed:
\begin{itemize}
\item[A)] if the affinity towards the interacting partner is below some
threshold, the individual tends to crystallize his/her actual
opinion, while for higher values of affinity he/she will change
opinion in the direction of the partner's one;
\item[B)] if the opinion difference between the two interacting agents
is below a critical value, then each one will increase his/her affinity
towards the partner, otherwise the affinity scores will decrease.
\end{itemize}
These two ways of acting are modulated by external factors, as for example the possibility
of interacting given by the social system, and especially by internal ones, such as
the \textit{openness of mind} and the
\textit{confidence}.
The openness of mind is the limit of
permissiveness that an individual introduces interacting with other
people, and allows to ignore the perception of incompatibilities
existing between oneself and the others; consequently, it makes possible to interact
with individuals having very distant opinions.
On the other hand, the confidence is the
minimal reputation an individual requires to a stranger to accept instances from him/her.
In practice, I will be more available to uniform my opinions with the opinions of someone
with a large affinity with myself.
Moreover, affinity acts as a long term memory in
which individuals can store information useful to solve similar
future situations.

By formalizing agents in such a way, we obtain a dynamical
population where interacting agents share their opinions by trying
to maintain an acceptable level of dissonance. The asymptotic states
of such system are either a global consensus (i.e. into an
hypothetical opinion space, a mono-clustered state) or a social
fragmentation (i.e. crystallization of no longer interacting
clusters of opinion). Of course, in the vision of the $PM$s, social fragmentation
has to be considered a dangerous state, since once obliged to
interact, the low level of mutual affinity and the differences in
opinion, may lead to strong social contrasts between these agents.
For this reasons, the goal of the $PM$s can be translated into a
reduction of the social fragmentation, namely into a reduction of
opinion distances among agents into the opinion space.

The aim of this paper is to present two possible models of $PM$
behavior. In the first case, we emphasize principally the skill of
interacting and negotiating with people along large opinion
distances. We label these $PM$s as ``\textit{diplomats}'' and we tag
their most prominent characteristic as a larger openness of mind. Classical
examples are actual diplomats, transactors, intermediaries, etc. On
the other hand, we consider as another fundamental attribute of a $PM$ his/her
reputation. Hence, we label this $PM$ figure as an ``\textit{auctoritas}'' (``authority''),
which is characterized by an established good reputation and the aptitude to
influence the society by their prestige. For example, we can set in this
category Mahatma Gandhi and Nelson Mandela.

Targets of this work are to obtain a mathematical representation of
both $PM$ figures and to investigate by means of numerical simulations
how they can affect a formalized social system of \textit{normal}
agents in order to reach a widespread social consensus.

\ 

The paper is organized as follows. The next section is dedicated to
describe the model. Then, in third section we present the numerical
results. Fourth and fifth sections are devoted to analytical
considerations and theoretical discussion, respectively. Finally, in
the last section we will sum up and talk about future perspectives.

\

\section{The Model}
\label{model}

We approach the problem by means of the tools and methods of
Sociophysics~\cite{Cas,Sen}.
In particular, the adopted model, without $PM$s, has already
been studied in previous studies~\cite{bag07,4,5}: indeed, we
are going to refine it in the present paper.
Therefore, we briefly recall its main features. The model is characterized by a
continuous opinion and a random binary encounter dynamics, as in
the Deffuant Model~\cite{Def1,Def2}, which ours is inspired to (at least the part
concerning the evolution of the opinions). We
consider a system made up of $N$ autonomous agents, the
individuals, each one identified by the index $i=1,\dots,N$ and characterized
by the two (constant) parameters $\Delta O^i_c$ and $\alpha^i_c$, which are the
openness of mind and the confidence\cite{nota_conf}, respectively. Moreover, each agent
$i$ is in general described by the two internal variables $O_i$, its opinion, and
$\alpha_{ij}$ ($j=1,\dots,N;\ j\neq i$), its affinity towards the other agents, which
vary in time. All $\Delta O^i_c$, $\alpha^i_c$, $O_i$ and $\alpha_{ij}$ are real numbers
ranging in the interval $[0, 1]$.
The internal variables evolve self-consistently during time evolution.

More precisely, let us consider an agent $i$ interacting with another agent $j$: 
then, the opinion $O_i$ and the affinity $\alpha_{ij}$ ({\it i.e.}, the affinity
$i$ feels towards $j$) are updated as follows~\cite{bag07,4,5}:

\begin{eqnarray}
O_i^{t+1} &=& O_i^{t} - \mu \ \Delta O_{ij}^{t} \ \Gamma_1(\alpha^t_{ij}) \label{opinion} \\
\alpha_{ij}^{t+1} &=& \alpha_{ij}^{t} + \alpha_{ij}^{t} \
[1-\alpha_{ij}^{t}] \ \Gamma_2 (\Delta O^t_{ij}) \label{alpha}
\end{eqnarray}

\noi where the activating functions $\Gamma_1$ and $\Gamma_2$ read, respectively:

\begin{eqnarray}
\Gamma_1(\alpha^t_{ij}) &=& \Theta(\alpha_{ij}^{t}-\alpha^i_c) \\
\Gamma_2(\Delta O^t_{ij}) &=& 1-2\ \Theta(|\Delta O_{ij}^{t}|
- \Delta O^i_c)
\end{eqnarray}

\noi being $\Delta O^t_{ij}=O_i^t-O_j^t$ the difference at time $t$ between the two
opinion values of the interacting partners, $\mu$ a convergence
parameter and $\Theta(\cdot)$ is the Heaviside step function.
In practice, an agent $i$ interacting with another agent $j$ changes its opinion
only if its affinity towards $j$ is larger than its own confidence $\alpha_c^i$:
in that case, the opinion updates according to the already mentioned Deffuant rule.
Analogously,
agent $i$ evolves the affinity towards $j$ only if their opinions differ less than
$i$'s openness of mind. If this is the case, the updating of
$\alpha_{ij}^t$ is not linear: the logistic term keeps the affinity in the
interval $[0,1]$; moreover, it maximizes the change in affinity for
pairs with $\alpha_{ij}\simeq0.5$, corresponding to agents which have
not come often in contact. Couples with $\alpha_{ij}\simeq1$ (resp. 0)
have already formed their mind and, as expected, behave
more conservatively. Anyway, a more thorough justification of these rules,
also from a psycho-social point of view, can be found in reference~\cite{bag07}.

At each elementary time step the two interacting agents are selected as follows:
the agent $i$ is drawn with uniformly distributed probability from the population,
whilst agent $j$ is the one which minimizes the social metric
\begin{equation}
D_{ij}^t = d_{ij}^t + \eta (0,\sigma) \label{social_metric} \ , 
\end{equation}

\noi composed by the two terms, respectively the \textit{social
distance}

\begin{equation}
d_{ij}^t = \Delta O_{ij}^{t} (1-\alpha_{ij}^{t}) \qquad j=1,...,N
\qquad j \ne i \label{social_distance}
\end{equation}

\noi and the gaussian noise ($\eta$) with mean value zero and variance
$\sigma$ (which is also called \textit{social temperature}~\cite{2}), modulating the
mixing degree in the population. The above formulas mean that two
agents are more likely to interact when their opinions have a small
difference and/or their affinity is larger, but, due to the social temperature,
it is always possible an interaction between two individuals with very different
opinions or very small reciprocal affinity. More precisely, in absence of social
noise ($\sigma=0$), an agent will surely interact with the stranger which minimizes
the social distance $d_{ij}^t$, with $\sigma\rightarrow+\infty$ the interactions
are completely random, for intermediate values most matches will be between
individuals at short social distance, with few long-range interactions.
A time unit is made up of $N$ single elementary time steps (Montecarlo steps).

Being the ultimate goal of $PM$s the reduction of social
fragmentation, both \textit{diplomats} and \textit{auctoritates}
will act in this direction, but via different courses of action.
\textit{Diplomats} are assumed to have a larger $\Delta O_c$ then
\textit{normal} agents and consequently they can interact in the
opinion space with far away agents. According to Eq. \ref{opinion},
this way of acting will lead to an increase of the individuals
affinity towards \textit{diplomats}. On the other hand,
\textit{auctoritates} are assumed to employ their notoriety; this is
translated in our model by imposing that all agents have a larger
affinity value towards them, directly promoting the convergence into
opinion space.

\

\section{Numerical simulations}
\label{numsec}

Simulations are performed with following parameters. $N$ is fixed
once for all to $100$, including $PM$s. The social temperature $\sigma$, the affinity threshold $\alpha_c$ and the convergence
parameter $\mu$ are fixed once for all, respectively at $0.003$,
$0.5$ and $0.5$. \textit{Normal} agents have a $\Delta O_c = 0.2$,
while for \textit{diplomats} $\Delta O_c = 0.5$. Entries in the
affinity matrix $\alpha$ are initialized between \textit{normal}
agents with uniformly distributed probability in $[0, 0.5]$, while
entries corresponding to \textit{normal} agents towards
\textit{auctoritates} are set at $0.75$.
We have chosen the above values of the parameters as the most reasonable and
conservative possible, in order not to have an unbalanced system (agents
not too mind-opened nor too mind-closed, affinity distribution not too narrow nor
too broad, etc.); moreover, the chosen value for $\sigma$ allows the existence of
interactions among socially distant agents maintaining higher probabilities for
matches between socially closer individuals. Anyway, we have also verified that
the results we are going to present in this section are rather robust by varying
such values: unless extreme values are chosen, the system behavior qualitatively
does not change.

We have considered both the fraction of $PM$s over the entire
population and their distribution in the opinion space as the
relevant control parameters, hereby measuring the mean number of
survived clusters at the equilibrium over 100 runs. The range of
employed $PM$s is from $5\%$ to $50\%$ in steps of $5\%$.
We remark that, so as formalized, the increase of fraction of $PM$s can
corresponds respectively either to a fixed number of $PM$s having to
do with smaller group, or to a population having a higher mean
$\Delta O_c$ (\textit{diplomats}).

Runs are stopped when the system converge to an equilibrium
asymptotic state. We define such a state is reached when the
affinity matrix will no longer change. We know that for communities
larger than $20$ agents, the system converge with respect to the
opinion before than respect to the affinity~\cite{5}. Hence, when
affinity reaches a state where it no longer evolves, the whole
system, i.e. also the opinion, will freeze. Such asymptotic state
will be characterized by the number of clusters in the opinion
dimension.

\textbf{Scenarios}. The behavior of the two $PM$s figures are
separately studied in a starting system which entries of opinion
vector $O$ are initialized uniformly spaced in $[0,1]$.
\textit{Diplomats} are distributed along the opinion space by
substituting them to the already initialized \textit{normal} agents
and according with the following modalities. In the
``\textit{uniform}'' distribution \textit{diplomats} are spread
along the opinion space with uniformly distributed probability; in
the ``\textit{gaussian}'' one with a gaussian distribution (mean
$0.5$, standard deviation $0.2$); in the ``\textit{bimodal}''
distribution they are inserted with a bimodal distribution, that is,
half of them with initial opinion equal $0.25$ and half equal to $0.75$.

The same opinion vector initialization and strategy distribution
are used for \textit{auctoritates}, with the addition of a
``\textit{delta}'' strategy in which all \textit{auctoritates} are
grouped around the center of the opinion space, namely around $0.5$.

\textbf{The ``\textit{two opposing factions}'' case -} Hereby we
propose an application of the model. We consider a starting opinion
space in which agents are divided into two large clusters, such that
their respective opinion distances are larger than the opinion
threshold of any single agent (``bi-clustered system''). In such a way, there is no
possibility of interaction between agents belonging to the two
different groups. Nevertheless, \textit{diplomats} are able to
interact with both factions because of their large openness of mind,
while \textit{auctoritates} can attract individuals because of their
high reputation. We thus compare the two different courses of
action.
In order to evaluate the effectiveness of their action, we tested every
configuration with a different density of mediators, from zero up to $50\%$
(of course, such a high presence of mediators does not take place in the real
world, anyway it is useful to reach it for a better theoretical understanding of
their role).

\

\section{Results}
\label{ress}

%%%%%%%%%%%%%%%%%%%%%%%%%%%%%%%%%%%%%%%%%%%%%%%%%%%%
\begin{figure}[ht!]
\centering
\includegraphics[width=16.0cm]{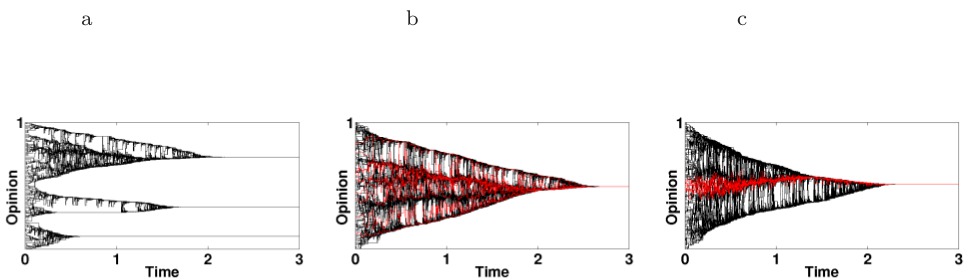}
\caption{Typical opinion trajectories. Each time
step are $10^{4}$ interactions. a) \textit{Normal} agents (see ref.~\cite{bag07});
b) \textit{Normal} agents (black) and \textit{diplomats}
(red); c) \textit{Normal} agents (black) and \textit{auctoritates}
(red).}\label{fig:fig1}
\end{figure}
%%%%%%%%%%%%%%%%%%%%%%%%%%%%%%%%%%%%%%%%%%%%%%%%%%%%%

Figure \ref{fig:fig1} shows typical trajectories into the opinion
space of a system of \textit{normal} agents (1a), a system
influenced by \textit{diplomats} (1b) and a system influenced by
\textit{auctoritates} (1c), respectively. While the system of
\textit{normal} agents quickly converge to a fragmented asymptotic
state, the insertion of $PM$s increases the convergence time needed
as so as the chances of obtaining a mono-clustered state. We remark
the different courses of action of the two $PM$s. Because of the
great $\Delta O_c$ value, \textit{diplomat} increases affinity
towards neighborhood, approaches partner and inclines it towards
its own opinion. Agents inside the opinion bounds of
\textit{diplomat} have a larger probability of collapse in the same
final position, and the \textit{diplomat} has the possibility to
explore the entire opinion space. On the other hand,
an \textit{auctoritas} tends to reach the equilibrium with the same
opinion value with respect to the initial condition. In this latter
case, the affinities of \textit{normal} agents towards
\textit{auctoritates} trigger the convergence dynamics to
monocluster.

%%%%%%%%%%%%%%%%%%%%%%%%%%%%%%%%%%%%%%%%%%%%%%%%%%%%%
\begin{figure}[h!]
\centering
\includegraphics[width=15.0cm]{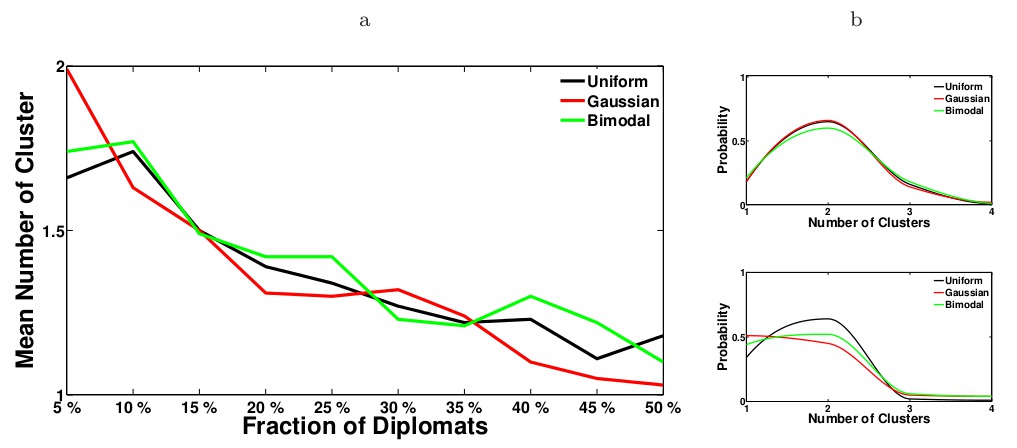}
\caption{Behavior of a system modulated by
\textit{diplomats}, for three different initial distributions.
a) Mean number of survived clusters at the equilibrium as a function
of the fraction of \textit{diplomats}. b) Probability of having $N$ clusters
at the equilibrium in a single run, 5\% of \textit{diplomats} (upper
figure), 50\% of \textit{diplomats} (lower); for sake of clearness,
the histograms are interpolated by ninth degree
polynomials (naturally, the effectively measured probabilities are in
correspondence of integer values of the abscissas).}\label{fig:fig2}
\end{figure}
%%%%%%%%%%%%%%%%%%%%%%%%%%%%%%%%%%%%%%%%%%%%%%%%%%%%%

Figure \ref{fig:fig2} resumes results relative to
\textit{diplomats}. The insertion of \textit{diplomats} reduces the
mean degree of fragmentation at equilibrium. Moreover, this
reduction is linear and positively correlate with the fraction of
employed \textit{diplomats}. Although the three distribution
strategies have similar trends (Fig. 2a), by augmenting the fraction
of \textit{diplomats}, the \textit{gaussian} one tends to reach the
greater number of mono-clusters at equilibrium (Fig. 2b, lower).

%%%%%%%%%%%%%%%%%%%%%%%%%%%%%%%%%%%%%%%%%%%%%%%%%%%%%
\begin{figure}[h!]
\centering
\includegraphics[width=15.0cm]{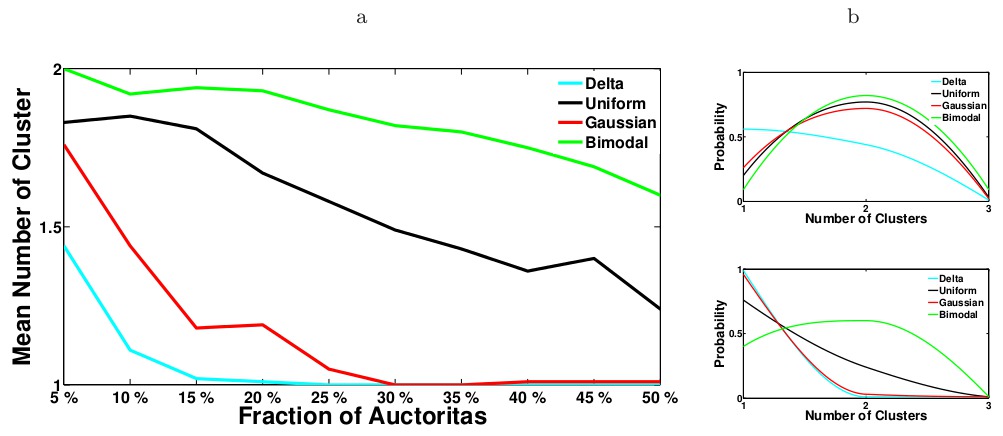}
\caption{Behavior of a system modulated by
\textit{auctoritates}, for four different initial distributions.
a) Mean number of survived clusters at the equilibrium as a function
of the fraction of \textit{auctoritates}. b) Probability of having $N$
clusters at the equilibrium in a single run, 5\% of \textit{auctoritates}
(upper figure), 50\% of \textit{auctoritates} (lower); for sake of
clearness, the histograms are interpolated by ninth degree polynomials
(naturally, the effectively measured probabilities are in
correspondence of integer values of the abscissas).}\label{fig:fig3}
\end{figure}
%%%%%%%%%%%%%%%%%%%%%%%%%%%%%%%%%%%%%%%%%%%%%%%%%%%%%

Figure \ref{fig:fig3} resumes results relative to
\textit{auctoritates}. Once more the insertion of $PM$s reduces the
mean degree of fragmentation at equilibrium, but hereby the adopted
distribution strategies substantially influence results of
simulations (Fig. 3a). By varying the employed fractions of
\textit{auctoritates}, \textit{gaussian} and, mainly, \textit{delta}
distributions show best trends in terms of convergence to a
mono-cluster state. The \textit{bimodal} distribution tends to
converge to a bi-clustered state (Fig. 3b, lower).

%%%%%%%%%%%%%%%%%%%%%%%%%%%%%%%%%%%%%%%%%%%%%%%%%%%%%
\begin{figure}[ht]
\centering
\includegraphics[width=16.0cm]{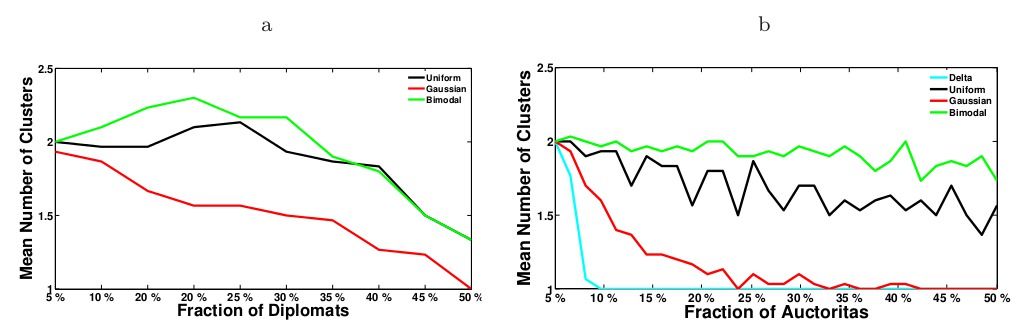}
\caption{Mean number of survived clusters in the final configuration as a
function of the fraction of $PM$s, for different
initial distributions of them, in case of an initially bi-clustered system. a) Acting
\textit{diplomats}; b) Acting
\textit{auctoritates}.}\label{fig:fig4}
\end{figure}
%%%%%%%%%%%%%%%%%%%%%%%%%%%%%%%%%%%%%%%%%%%%%%%%%%%%%

Figure \ref{fig:fig4} shows results of insertion of $PM$s into a
bi-clustered starting population (that is, a population where initially half population
has opinion equal to 0.25 and the remaining half equal to 0.75); previous results are confirmed.
\textit{Diplomats} become efficacious only for higher fractions of
employment and mainly with a \textit{gaussian} distribution.
\textit{Auctoritates}, spread with either a \textit{gaussian} or,
above all, a \textit{delta} strategy, assure the convergence to a
mono-clustered asymptotic state since lower fractions of employment.

\begin{figure}[h!]
\centering
\includegraphics[width=9.0cm]{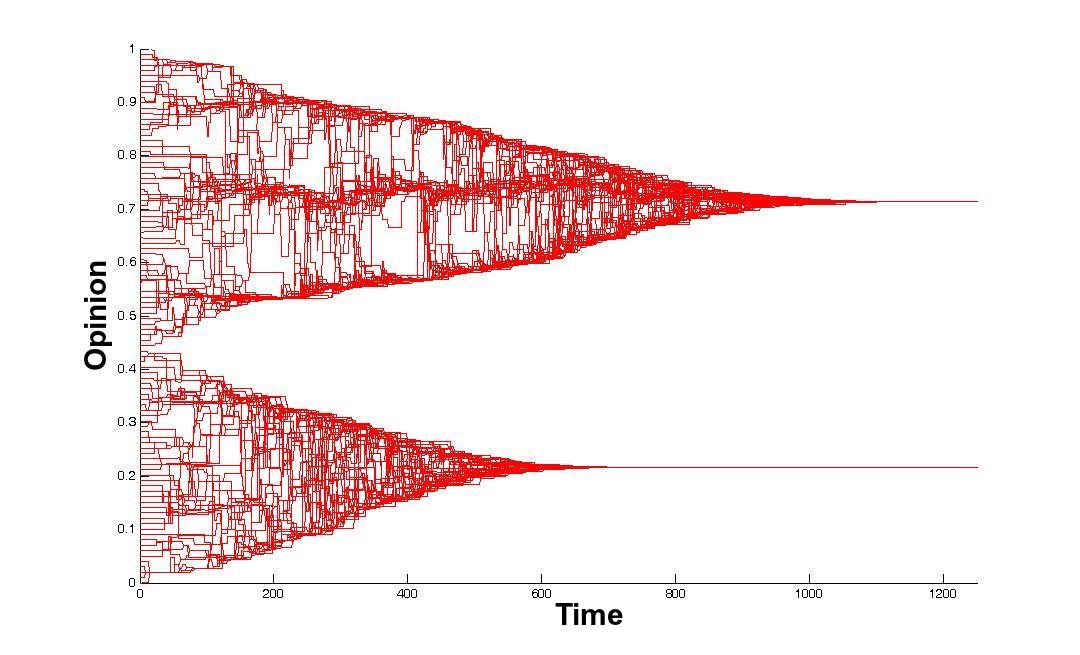}
\caption{
Opinion trajectories for a system of $N=100$ individuals, with $5\%$ of
{\it diplomats} initially uniformly distributed. In this realization, the
system has ended up with two final clusters.}
\label{op2cl}
\end{figure}

\begin{figure}%[h]
\begin{centering}
\includegraphics[width=6.1cm]{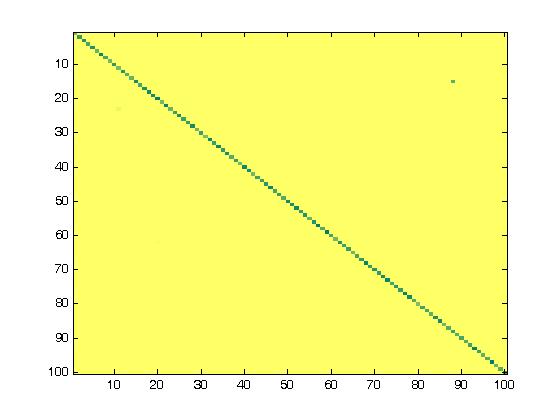}
\includegraphics[width=6.1cm]{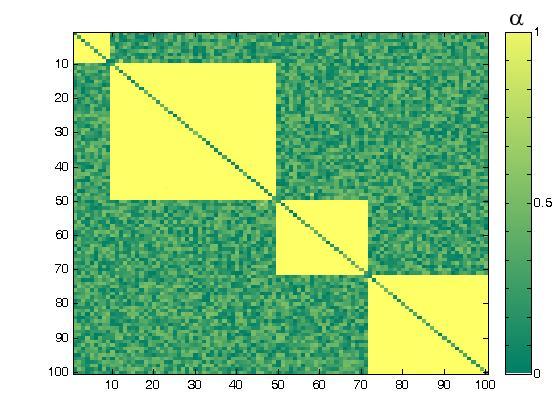}
\end{centering}
\caption{
Color representation of the final affinity matrices for systems of $N=100$
individuals and $10\%$ of initially delta-distributed {\it auctoritates}
(left, final monocluster), and with no PMs at all (right, four final clusters).
The affinities of the agents towards themselves, irrelevant for the dynamics,
were set to $0$.}
\label{fin_al}
\end{figure}

{\bf Opinion and affinity final distributions -} Concerning the configuration
of the opinions and affinities at the end of the dynamics, we verified that
the former distribute so that the surviving ones are all equidistant, that
is, if there is a monocluster the unique final opinion will be around $0.5$
(see Fig.~\ref{fig:fig1}), if there are two cluster the survived opinions
will be close to $0.25$ and $0.75$ (see Fig.~\ref{op2cl}), and so on.
On the other hand, the final affinities distribute in the simplest way:
agents belonging to the same cluster ({\it i.e.}, share the same final
opinion) will have affinity equal to 1 towards each other, and practically
to 0 if instead they end up with different opinion, as shown in Fig.~\ref{fin_al}.

%%%%%%%%%%%%%%%%%%%%%%%%%%%%%%%%%%%%%%%%%%%%%%%%%%%%%
%%%%%%%%%%%%%%%%%%%%%%%%%%%%%%%%%%%%%%%%%%%%%%%%%%%%%

%%%%%%%%%%%%%%%%%%%%%%%%%%%%%%%%%%%%%%%%%%%%%%%%%%%%%
%%%%%%%%%%%%%%%%%%%%%%%%%%%%%%%%%%%%%%%%%%%%%%%%%%%%%

\ 

\section{Theoretical analysis}
\label{Tan}

The dynamics of our model is complex and highly non-linear, so that it would
be quite hard to get analytically the simulation results by means of a complete
theoretical calculation. Anyway, it is possible to provide a qualitative
justification of the behavior of the model. Let us start by considering the case of absence of any PM, that is, when every agent has the same openness of mind $\Delta O_c$ and all the affinities $\alpha_{ij}$ have the same (uniform) distribution $\forall i,j$. In Figure~\ref{fig:fig1}a an example of how dynamics takes place in this case is depicted. We can utilize a simple mean-field treatment to describe the dynamics. Therefore, said $P(x,t) dx$ the fraction of agents having an opinion in the range $[x,x+dx]$, and having in mind the definitions given in Section~\ref{model}, the rate equation of the distribution $P(x,t)$ is given by

\be 
\label{mf1}
\frac{\partial P(x,t)}{\partial t}=\int_0^1dO\int_{|O-O'|<\frac{\sigma\Delta O_c}{1-\langle\alpha_{ij}\rangle}} dO'\ P(O,t)P(O',t)\left[\delta(x-O+\mu(O-O'))-\delta(x-O)\right] \ , 
\ee

\noi where we remind that $\sigma$ is the social temperature. Analogously, it is also  possible to write down an equation for the evolution of the affinities:

\be 
\label{mf2}
\frac{d\alpha_{ij}}{dt}=\Gamma(t)\ \alpha_{ij}[1-\alpha_{ij}(t)] \ , 
\ee

\noi where $\Gamma(t)=\mbox{sgn}\left(|\Delta O_{ij}(t)-\Delta O_c|\right)$. Now, let us focus on Eq.~(\ref{mf1}): it is formally completely analogous to the rate equation of the compromise model (CM) defined in~\cite{Ben}. In the CM two individuals interact only if their opinions differ less than 1, but the (continuous) opinions are defined up to a certain maximum value $O_M$. It results that the system ends up to final consensus only if $O_M<1$: in practice, the system orders if the maximum opinion difference is not larger than the threshold for an interaction to take place. In our model $\max|O-O'|$ is of course 1,
while the threshold is, in the mean-field approximation we are utilizing,
$\sigma\Delta O_c/(1-\langle\alpha_{ij}\rangle)$, which defines the integration
interval of the integral in $dO'$ in Eq.~(\ref{mf1}). Therefore,
proceeding in the same way of reference~\cite{Ben}, it becomes crucial the quantity

\be 
\label{delta}
\Delta\equiv\frac{\sigma\Delta O_c}{1-\langle\alpha_{ij}\rangle} \ , 
\ee

\noi such that if it is larger than 1 (the maximum difference possible between two opinions), the system reaches consensus, otherwise it remains disordered. Basically, in order to have final consensus it must hold

\be 
\label{delta_rel}
\Delta\geq\Delta_c=1 \ . 
\ee

\noi The existence of this transition is confirmed in Ref.~\cite{bag07}, even though this mean-field approximation does not catch its exact behavior. On the other hand, in Figure~\ref{fig:fig1}a we show the time evolution of a system with $\sigma=0.003$, $\Delta O_c=0.2$ and $\langle\alpha_{ij}\rangle=0.25$ (see Ref.~\cite{bag07} again), that is with $\Delta=0.0008\ll\Delta_c$, and actually consensus is not reached, compatibly with the considerations stated above.
We highlight the role of the social temperature: if $\sigma\rightarrow0^+$,
the probability of reaching consensus goes to zero (rigorously, in the limit
of infinite system), because a non-zero social temperature makes two far away agents interact.

Despite the roughness of the previous calculations, this approach allows us to reckon at least qualitatively the effect of the presence of the peace mediators on the ultimate fate of the system. As a matter of fact, \textit{auctoritates} increase the average $\langle\alpha_{ij}\rangle$, and in a similar way \textit{diplomats} increase the average $\langle\Delta O_c^i\rangle$, which substitutes the simple $\Delta O_c$ in Eqs.~(\ref{mf1}) and~(\ref{delta}): therefore, both make the quantity $\Delta$ of the system larger, enhancing the reaching of final consensus, as confirmed by the numerical results presented in Section~\ref{ress}.

\ 

\section{Discussion}

In the above sections we showed, both numerically and theoretically,
that the presence of the $PM$s helps the system to reach more easily
the final consensus, with respect to the model without $PMs$, treated
in reference~\cite{bag07}. It results
that these two kinds of special agents act differently and have different
effects. More precisely, \textit{diplomats}' effects are quite independent from their distribution
throughout the population, as shown in Figure~\ref{fig:fig2}a, whilst \textit{auctoritates}'
action is clearly sensitive to their dislocation in the opinion space (see
Figure~\ref{fig:fig3}a). Moreover,
when the \textit{auctoritates} are effective in favor of consensus, a smaller number of them
is required with respect to the case with \textit{diplomats}.
In short, the system shows the best response in terms of final consensus when few
\textit{auctoritates} are put just in the middle of the social space of the population
({\it i.e.}, with opinion equal to the average of the system, in our case 0.5).
These results are not in contrast with some real world features, in particular
with the fact that mediators like \textit{auctoritates} are less common but more effective
than \textit{diplomat}-like agents. Indeed, while the main characteristic of a \textit{diplomat},
a larger openness of mind, depends only on the individual itself, acquiring
authority to the others' eyes is much more difficult and does depend in
general on the behavior towards other people. On the other hand, the personal
prestige, once obtained, is certainly more incisive in order to persuade other
individuals. For this same reason, when the \textit{auctoritates} differ very much in
opinion, they find much harder to drive the system to the consensus: as we can
see in the real world, if several charismatic figures push the people towards
opposite positions, usually the whole population is not able to get a general
agreement.

Considering now a more theoretical point of view, as illustrated in
Section~\ref{Tan}, the action of the $PM$s is effective because practically
they help to reduce the average distance among the individuals in the opinion
space. Actually, in their absence, unless all the individuals have a very
high openness of mind and/or affinities towards each other, the only way to
increase the probability to reduce conflict and reach consensus is to increase
the social temperature, helping people to interact despite their differences,
as illustrated in our previous works~\cite{bag07,5}.
Anyway, the social temperature must be regarded as an intrinsic property
of the environment in which agents find themselves to act, and it would be
difficult for whatever institution to change it in practice: on the contrary,
it is clearly possible to send to the population in conflict some Peace
Mediators.

Otherwise, we can wonder how to reach similar results when the dynamics
of the agents is not as defined here. For instance, let us think to different types
of continuous opinion models, in particular the LCCC model~\cite{lal10} and
similar ones~\cite{sen12}. In these models, agents exchange their opinions,
but depending on the value of a parameter called \textit{conviction}, that is,
the ``inertia'' an individual opposes to mind changing, the population can reach
distinct final states. When the individuals have a high conviction, the system
is driven to a configuration where the individuals have extreme opinions, giving
birth to two totally contrasting factions. In such a case,
mediators should simply aim to reduce the agents' convictions in order to reduce
conflicts. With respect to these models, ours is more complex since it is defined
by more parameters, but has the advantage to represent at least partially
the internal dynamics of the agents' minds~\cite{gia15}.

\ 

\section{Conclusions and future perspectives}

In this paper we propose an application of the model of continuous
opinion dynamics already introduced in~\cite{bag07,4,5}, by inserting
two figures of Peace Mediators, one by one either \textit{diplomats} or
\textit{auctoritates} respectively, into a population of
\textit{normal} agents. We describe the behavior of the system in
terms of opinion convergence and mean degree of fragmentation for
different fraction of employed $PM$s, also in reference to a more
likely situation, namely the case ``\textit{two opposing
factions}''.

The typical \textit{modus operandi} of \textit{diplomats} becomes
more effective by inserting many of them. By referring to what we said
in section~\ref{numsec},
both the insertion of few \textit{diplomats} into groups of small
size and the increase of the mean $\Delta O_c$ value of the
population would lead to the same result. On the other hand, the
promotion of few \textit{auctoritates}, but in suitable positions,
can assure the convergence to a widespread consensus into
populations of any sort, as pointed out Sections~\ref{ress} and~\ref{Tan}.

The combined efforts of both the two kinds of $PM$s remain to be further
investigated: in this paper we considered them always acting separately
because our aim was the understanding of their disentangled effects, in
order to evaluate more precisely their role and features. 
Also the effects of the population size, the time needed by
such figures in order to reach the global consensus and the role of hypothetical
powerful neighboring (which could have interest in fostering or hindering consensus)
are still to be better understood. Finally, here we set the $PM$s as naturally different
from the normal agents, but it could be worth to understand if they can emerge from
a suitable evolutionary framework. To address all these issues, deeper studies,
both of analytical and experimental nature, are needed in the next future.

\ 

\ 

\section*{Acknowledgements}

Work supported by Commission (FP7-ICT-2013-10) Proposal No. 611299 SciCafe 2.0,
by H2020 FETPROACT-GSS CIMPLEX Grant No. 641191,
and by project CLARA (CLoud plAtform and smart underground imaging for natural Risk Assessment), funded by the Italian Ministry of Education and
Research (PON 2007-2013: Smart Cities and Communities and Social Innovation; Asse e Obiettivo: Asse II - Azione Integrata per la Societ\`a dell'Informazione; Ambito: Sicurezza del territorio).

\ 

\ 

%% The Appendices part is started with the command \appendix;
%% appendix sections are then done as normal sections
%% \appendix

\section*{References}

%% \label{}

%% References
%%
%% Following citation commands can be used in the body text:
%% Usage of \cite is as follows:
%%   \cite{key}          ==>>  [#]
%%   \cite[chap. 2]{key} ==>>  [#, chap. 2]
%%   \citet{key}         ==>>  Author [#]

%% References with bibTeX database:

%% \bibliographystyle{model1-num-names}
%% \bibliography{vcbg_2016.bib}

%merlin.mbs apsrev4-1.bst 2010-07-25 4.21a (PWD, AO, DPC) hacked
%Control: key (0)
%Control: author (8) initials jnrlst
%Control: editor formatted (1) identically to author
%Control: production of article title (-1) disabled
%Control: page (0) single
%Control: year (1) truncated
%Control: production of eprint (0) enabled
\begin{thebibliography}{0}%
\makeatletter
\providecommand \@ifxundefined [1]{%
 \@ifx{#1\undefined}
}%
\providecommand \@ifnum [1]{%
 \ifnum #1\expandafter \@firstoftwo
 \else \expandafter \@secondoftwo
 \fi
}%
\providecommand \@ifx [1]{%
 \ifx #1\expandafter \@firstoftwo
 \else \expandafter \@secondoftwo
 \fi
}%
\providecommand \natexlab [1]{#1}%
\providecommand \enquote  [1]{``#1''}%
\providecommand \bibnamefont  [1]{#1}%
\providecommand \bibfnamefont [1]{#1}%
\providecommand \citenamefont [1]{#1}%
\providecommand \href@noop [0]{\@secondoftwo}%
\providecommand \href [0]{\begingroup \@sanitize@url \@href}%
\providecommand \@href[1]{\@@startlink{#1}\@@href}%
\providecommand \@@href[1]{\endgroup#1\@@endlink}%
\providecommand \@sanitize@url [0]{\catcode `\\12\catcode `\$12\catcode
  `\&12\catcode `\#12\catcode `\^12\catcode `\_12\catcode `\%12\relax}%
\providecommand \@@startlink[1]{}%
\providecommand \@@endlink[0]{}%
\providecommand \url  [0]{\begingroup\@sanitize@url \@url }%
\providecommand \@url [1]{\endgroup\@href {#1}{\urlprefix }}%
\providecommand \urlprefix  [0]{URL }%
\providecommand \Eprint [0]{\href }%
\providecommand \doibase [0]{http://dx.doi.org/}%
\providecommand \selectlanguage [0]{\@gobble}%
\providecommand \bibinfo  [0]{\@secondoftwo}%
\providecommand \bibfield  [0]{\@secondoftwo}%
\providecommand \translation [1]{[#1]}%
\providecommand \BibitemOpen [0]{}%
\providecommand \bibitemStop [0]{}%
\providecommand \bibitemNoStop [0]{.\EOS\space}%
\providecommand \EOS [0]{\spacefactor3000\relax}%
\providecommand \BibitemShut  [1]{\csname bibitem#1\endcsname}%
\let\auto@bib@innerbib\@empty
%</preamble>
\end{thebibliography}%


\begin{thebibliography}{20}

\bibitem{fes58}
L. Festinger. {\it A Theory of Cognitive Dissonance}. Evanston
(1958).

\bibitem{bag07}
F. Bagnoli, T. Carletti, D. Fanelli, A. Guarino and A. Guazzini, {\it Phys. Rev. E}, {\bf 76}, 066105 (2007).

\bibitem{2}
R. E. Nisbett and L. Ross. {\it Human Inferences: Strategies and shortcomings
of social Judgement}. Englewood Cliffs, Prentice-Hall (1980).

\bibitem{Cas}
C. Castellano, S. Fortunato and V. Loreto, {\it Rev. Mod. Phys.}, {\bf 81}, 591 (2009).

\bibitem{Sen}
P. Sen and B. K. Chakrabarti. {\it Sociophysics - An Introduction}. Oxford (2014).

\bibitem{4}
T. Carletti, D. Fanelli, A. Guarino and A. Guazzini. {\it Meet Discuss and trust each
other: large versus small groups}. Proceedings in {\it WIVACE2008 Workshop Italiano Vita Artificiale e Computazione Evolutiva} (2008).

\bibitem{5}
T. Carletti, D. Fanelli, A. Guarino, F. Bagnoli and A. Guazzini, {\it Eur. Phys. B}, {\bf 64}, 2, 285 (2008).

\bibitem{Def1}
G. Deffuant, D. Neau, F. Amblard and G. Weisbuch, {\it Adv. Compl. Syst.}, {\bf 3}, 87 (2000).

\bibitem{Def2}
G. Weisbuch, G. Deffuant, D. Neau and F. Amblard, {\it Interacting agents and continuous opinions
dynamics}, arXiv:cond-mat/0111494 [cond-mat.dis-nn] (2001).

\bibitem{nota_conf}
We highlight the fact that this \textit{confidence} here defined is a different concept from the parameter
with the same name utilized in the Deffuant Model.

\bibitem{Ben}
E. Ben-Naim, P. L. Krapivsky and S. Redner, {\it Phys. D}, {\bf 183}, 190 (2003).

\bibitem{lal10}
M. Lallouache,  A. S. Chakrabarti, A. Chakraborti and B. K. Chakrabarti,
{\it Phys. Rev. E}, {\bf 82} 056112 (2010).

\bibitem{sen12}
P. Sen, {\it Phys. Rev. E}, {\bf 86}, 016115 (2012).

\bibitem{gia15}
F. Giardini, D. Vilone and R. Conte, {\it Front. Phys.}, {\bf 3}, 00064 (2015).

%% \bibitem must have the following form:
%%   \bibitem{key}...
%%

% \bibitem{}

\end{thebibliography}

%% Authors are advised to submit their bibtex database files. They are
%% requested to list a bibtex style file in the manuscript if they do
%% not want to use model1-num-names.bst.

%% References without bibTeX database:

\end{document}